\documentclass[aps,prx,twocolumn,showpacs,superscriptaddress,groupedaddress]{revtex4-2}

\usepackage{graphicx}
\usepackage{dcolumn}
\usepackage{bm}

\usepackage{siunitx}
\usepackage{amsmath}
\usepackage{graphicx,amssymb}
\usepackage{tabularx}
\usepackage{booktabs}

\newcommand{\pp}{\mathcal{\partial}}

\newcommand{\pd}[2]{\frac{\partial #1}{\partial #2}}
\newcommand{\pdd}[2]{\frac{\partial^2 #1}{\partial #2 ^2}}
\newcommand{\tasd}{\overline{\delta^2(\Delta)}}
\newcommand{\msd}{\left< x^2(t)\right>}
\newcommand{\average}[1]{\left< #1 \right>}

\begin{document}

\preprint{APS/123-QED}

\title{Ergodicity breaking in area-restricted search of avian predators}




\author{Ohad Vilk$^{a,b,c}$}
\email{ohad.vilk@mail.huji.ac.il}
\author{Yotam Orchan$^{b, c}$}
\author{Motti Charter$^{b, c, d}$}
\author{Nadav Ganot$^{b, c}$}
\author{Sivan Toledo$^{c, e}$}
\author{Ran Nathan$^{b, c}$}
\email{ran.nathan@mail.huji.ac.il}
\author{Michael Assaf$^{a}$}
\email{michael.assaf@mail.huji.ac.il}

\affiliation{$^{a}$Racah Institute of Physics, The Hebrew University of Jerusalem, Jerusalem, Israel.}
\affiliation{$^{b}$Movement Ecology Lab, Department of Ecology, Evolution and Behavior, Alexander Silberman Institute of Life Sciences, Faculty of Science, The Hebrew University of Jerusalem, Jerusalem, Israel.}
\affiliation{$^{c}$Minerva Center for Movement Ecology, The Hebrew University of Jerusalem, Jerusalem, Israel.}
\affiliation{$^{d}$The Shamir Research Institute and Department of Geography and Environmental Studies, University of Haifa, 199 Aba Hushi Boulevard, Mount Carmel, Haifa, Israel.}
\affiliation{$^{e}$Blavatnik School of Computer Science, Tel-Aviv University, Israel.}


\begin{abstract}
Quantifying and comparing patterns of dynamical ecological systems require averaging over measurable quantities. For example, to infer variation in movement and behavior, metrics such as step length and velocity are averaged over large ensembles. Yet, in nonergodic systems such averaging is inconsistent; thus, identifying ergodicity breaking is essential in ecology. Using rich high-resolution movement datasets ($>\! 7 \times 10^7$ localizations) from 70 individuals and continuous-time random walk modeling, we find subdiffusive behavior and ergodicity breaking in the localized movement of three species of avian predators. Small-scale, within-patch movement was found to be qualitatively different, not inferrable and separated from large-scale inter-patch movement. Local search is characterized by long power-law-distributed waiting times with diverging mean, giving rise to ergodicity breaking in the form of considerable variability uniquely observed at this scale. This implies that wild animal movement is scale specific with no typical waiting time at the local scale.
\end{abstract}

\maketitle


\section{INTRODUCTION}

Movement of organisms is of key interest in many scientific fields, playing an essential role in a wide range of biological and ecological systems \cite{nathan2008movement}. Quantifying the patterns of such dynamical systems and elucidating their underlying mechanisms are typically based on analysis of key measurable quantities. A central challenge in the study of dynamical systems is to identify \textit{non}-ergodic processes, encompassing a discrepancy between long-time averaging over a time-dependent sample and an ensemble average across different samples \cite{barkai2012single}. Such a discrepancy can also entail ageing, indicating a tendency to decrease diffusivity over time~\citep{metzler2014anomalous}. Ergodicity breaking is thus of major interest in the study of diffusion processes~\cite{weigel2011ergodic, jeon2011vivo}; however, discerning between ergodic or nonergodic processes has been widely overlooked in many scientific disciplines~\citep{mangalam2021point}. 
In ecology, ergodicity breaking can result from variation among individuals in their internal states and traits, from  external environmental factors affecting individuals, or from various constraints shaping the interactions between these two types of effects~\citep{nathan2008movement, mendez2016stochastic}.

Another important aspect of movement is that it typically varies across spatiotemporal scales due to resource patchiness, seasonality or other environment features~\citep{nathan2008movement, levin1992problem, fryxell2008multiple, benhamou2014scales, martin2015coping, peron2019time}.
Inferring behavior from one scale to another can thus lead to nonrepresentative results~\citep{torney2018single, peron2019time}. In particular, real-life landscapes are typically heterogeneous, and animals routinely alternate between an \textit{extensive commuting} mode of movement between resource-rich patches, and an \textit{intensive searching} mode of area-restricted search (ARS) for prey within a local patch~\citep{bazazi2012intermittent, benhamou2014scales, torney2018single, riotte2020environmental}. 
Elucidating the drivers of this alternating behavior requires detailed information on animal movement both within and between patches at a high spatiotemporal resolution, from multiple individuals and over a sufficiently long time, which are difficult to obtain with standard wildlife tracking technologies~\citep{ran2022BigData}. Yet, averaging across conspecific individuals to compute scaling laws in animal movement defies the growing evidence of variability between individuals~\citep{macintosh2015edge, spiegel2015moving, campos2016variability, shaw2020causes}. In addition, many studies have distinguished between the commuting and searching (ARS) modes and recognized the hierarchical scale-related nature of space and habitat use \cite{johnson1980comparison, benhamou2014scales}. However, movement \textit{within} ARS, also known as fourth-order habitat selection \cite{johnson1980comparison} has seldom been analyzed due to data limitations \cite{ran2022BigData}, overlooking a critical scale of foraging essential for identifying the patterns and mechanisms underlying variation in movement behavior.

Ergodicity breaking is often studied using the celebrated framework of continuous-time random walks (CTRW)~\cite{montroll1965random}, defined in terms of the waiting times (WTs) between successive jumps, which is frequently used in physical, biological, and ecological systems~\citep{scher1975anomalous,shlesinger1982random, weigel2011ergodic, viswanathan2011physics, brockmann2006scaling}. An important subset of CTRW includes power-law-distributed WTs which can give rise to occasional long WTs and weak ergodicity breaking~\citep{metzler2000random, metzler2014anomalous}. 
Although similar continuous-time (and biased) random-walks have been extensively used in ecology, these have been applied mostly for statistical inference~\cite{johnson2008continuous, michelot2019state}, data fitting~\cite{fleming2014fine}, or simulation-based research, but seldom as a \textit{theory-based} tool~\cite{weigel2011ergodic, jeon2011vivo}, to investigate the core mechanisms that underlie movement processes in wild animals.
In ecology, long WTs have been empirically reported for ambush marine predators~\citep{wearmouth2014scaling}, free-ranging foragers such as seabirds~\citep{bartumeus2010fishery}, insects~\citep{bazazi2012intermittent}, cattle~\citep{zhao2016understanding}, and humans~\citep{brockmann2006scaling}. The underlying reasons for long WTs span from long rests~\citep{tilles2016random}, pauses to more effectively search for hidden prey and to organize attacks, or due to  interactions such as mating, territorial guarding and predator avoidance~\citep{obrien1990search, kramer2001behavioral}. Overall, animals exhibit a wide range of intermittent foraging behaviors, spanning from ambush with very long stops and short or no moves, through saltation (stop and go) alternating between intermediate stops and moves, to cruising (widely ranging) with constant movement and a few stops~\citep{obrien1990search, kramer2001behavioral}. 

In this study, we use high-resolution data from a new reverse-GPS wildlife tracking system~\citep{weiser2016characterizing, toledo2020cognitive} to characterize within-patch ARS movements of three species of avian predators: barn owls (\textit{Tyto alba}, hereafter ``owls''), black-winged kites (\textit{Elanus caeruleus}, ``kites''), and common kestrels (\textit{Falco tinnunculus}, ``kestrels''), all common residents in Israel. 
Focusing on ARS, 
we reveal that these predators display a fat-tailed distribution of WTs, giving rise to subdiffusion, ergodicity breaking and ageing at the local scale of a single patch. In contrast, commuting (non-local, between patches) flights are shown to be superdiffusive and ergodic. To do so, we employ the subdiffusive CTRW formalism, which is shown to adequately model movement within ARS. In particular, we show that the normalized time-averaged square displacement follows the Mittag-Leffer distribution as predicted for processes displaying subdiffusive CTRW~\cite{he2008random}. Finally, we find evidence of a behavioral switch that occurs between local ARS and non-local commuting, indicating scale-specific behavior of the studied species.


\section{Theoretical model} \label{sec:CTRW}
Ergodicity breaking is formally defined as a disparity between the mean square displacement (MSD) and  time-averaged square displacement 
(TASD)~\cite{metzler2014anomalous, mendez2016stochastic}.
The MSD is defined as the square displacement of an individual's position with respect to a reference position, averaged over an \textit{ensemble} of movement paths. In anomalous diffusion, the MSD satisfies
$
    \msd \sim t^\alpha,
$
where angular brackets denote ensemble averaging and $t$ is the measurement time. Here, the dynamics is superdiffusive for $\alpha > 1$ and subdiffusive for $\alpha < 1$, whereas $\alpha \to 1$ is the Brownian limit
\cite{metzler2000random}.
The TASD, also known as \textit{the semivariance function} used in several ecological studies (see, \textit{e.g.},~\cite{fleming2014fine}), is given by averaging over the square displacement performed in a time lag $\Delta$,
\begin{equation} \label{TASD}
\tasd = \frac{1}{t - \Delta}\int_0^{t - \Delta} [x(t' + \Delta) - x(t')]^2 dt',
\end{equation}
where an overline denotes time averaging. For simple Brownian motion (\textit{e.g.}, Pearson's RW~\citep{pearson1905problem}) and $\Delta \ll t$ one obtains $\tasd \sim \Delta \sim \left< x^2(\Delta)\right>$. Moreover, the TASD does not depend on the total measurement time $t$. In contrast, if the TASD and MSD scale differently, the underlying process is, by definition, nonergodic; that is, the \textit{ensemble} averaging is different from the \textit{time} averaging~\cite{metzler2000random}. 
Note that while not all subdiffusive processes display ergodicity breaking~\citep{mendez2016stochastic}, we show below that movement within ARS is both subdiffusive and nonergodic.
Note that, in many realistic cases, the TASD is a more convenient empirical measure than the MSD as the former provides robust statistics in the limit $\Delta \ll t$ and does not require a reference position or time, which is often arbitrary or unknown in ecological systems~\citep{metzler2014anomalous}.

To assess anomalous diffusion within ARS, we apply the CTRW formalism (see Appendix A for details), and define the WT, $\tau$, between successive jumps as a random variable drawn from probability distribution function $\psi(\tau)$. When the average WT $\average{\tau}$ diverges, the process displays subdiffusive dynamics and ergodicity breaking~\citep{metzler2000random, metzler2014anomalous}. We assume power-law-distributed WTs,
\begin{equation} \label{WT_PDF}
\psi(\tau) \sim \tau^{-(1+\alpha)},
\end{equation}
which, for $0< \alpha < 1 $, yield a diverging mean and ergodicity breaking. In  contrast, simple Brownian motion is generally a Poisson process with exponentially-distributed WTs, $\psi(\tau) = \tau_0^{-1} \exp(-\tau/\tau_0)$, where $\tau_0$ is the mean WT, resulting in ergodic dynamics.
Notably, in CTRW the jump length can also be taken as a random variable; we have numerically verified that our main results are independent of the jump length choice  (see Appendix A).

To quantify the subdiffusive process at hand, we compute the so-called \textit{averaged} TASD~\citep{metzler2014anomalous}, which can be shown to scale at long measurement times $\Delta \ll t$ as~\citep{burov2010aging}:
\begin{equation} \label{averagedtasd_def}
\average{\tasd} = 1/N \sum_{i = 1}^{N} \tasd\sim \left( \Delta/t \right)^{1- \alpha}.
\end{equation}
Here the TASD is averaged over an ensemble of $N$ samples (i.e., ARS segments); averaging is necessary  due to the irreproducible nature of the process (i.e., large diversity across trajectories). Moreover, the scaling is obtained by assuming that the movement is confined to a spatially bounded subregion (strictly speaking, due to an external confining potential, see Appendix~\ref{appendix:math} for mathematical details). Importantly, this scaling is significantly different than that expected for simple Brownian motion and most ergodic processes, where the TASD saturates upon interacting with the confinement.



\section{Methods}
\subsection{Data collection} \label{Sec:data}
Sixty owls were tracked in the Hula Valley, Israel (33.10N, 35.61E) between May and December 2018. Eighteen adults were tracked both during and subsequent post breeding and 42 fledglings were tracked for the first few months after fledging. For the adults, we used the hatching date to define the breeding season by assuming that the 90 days following hatching are within the season, as nestlings still depend on their parents~\citep{taylor2004barn}. Twenty-one kites were tracked in the Hula Valley between July 2019 and September 2020. Six were adults that actively bred during the tracking period, and the other 15 were fledglings. As kites can have multiple broods in a year, we defined their breeding season by directly observing their nests. Lastly, 15 kestrels were tracked in the Judean Plains, Israel (31.74N, 34.84E) between March and August 2019. Here, 11 were actively nesting, and the breeding season was defined by direct observations.

Individuals were tracked using ATLAS (Advanced Tracking and Localization of Animals in real-life Systems), an innovative reverse-GPS system that localizes extremely light-weight, low-cost tags~\citep{weiser2016characterizing}. Each ATLAS tag transmits a distinct radio signal which is detected by a network of base-stations distributed in the study area. Tag localization is computed using nanosecond-scale differences in signal time-of-arrival to each station, allowing for real-time tracking and alleviating the need to retrieve tags or have power-consuming remote-download capabilities~\citep{toledo2020cognitive, vilk2022phase}. The individuals tracking frequency was between 0.125 and 1 Hz. Localization errors are reported as a $2\times 2$  covariance matrix per localization. In this study we omit localizations with variance $> 50^2 \;m^2$, defined in terms of the trace over the covariance matrix. Furthermore, we filtered out days or nights in which many localizations are missing ($> 70\%$). Notably, in accordance with the typical error reported by the system ($\sigma \simeq5$ m)~\citep{weiser2016characterizing, beardsworth2021validating}, we assume $10$ m to be the noise limit in our measurements. While for many ARS the noise is practically much smaller, this is treated as an upper limit for any significant results. In total, our analyses incorporated high-quality data for 4,710 nights and $> 5\times 10^7$ localizations for 44 owls, 1,619 days and $> 10^7$ localizations for 16 kites, and 508 days and $> 9\times 10^6$ localizations for 10 kestrels, mostly during their respective breeding seasons.
We collected additional information for the adult birds: location of the nest, sex, breeding status, and brood size (when relevant). Only individuals with $> 15$ tracking days were included in the study: 14 adult and 30 fledgling owls, 6 adult and 10 fledgling kites, and 10 kestrels.
We limited the analyses to movement data collected during the activity hours (nights for the nocturnal owl and days for the other two diurnal species) and excluded data collected in proximity to the nest for breeding birds to focus on local search behavior (ARS).

\subsection{Statistical analysis} \label{stat_ana}
Localizations of all individual birds were segmented into ARS and commuting segments by detecting switching points in the data -- distinct points in which the bird switches between the two behaviors~\citep{benhamou2014scales}. 
ATLAS tracks were first segmented to exclude stops in or around the (known) nests of breeding adults. The remaining tracks were considered foraging excursions and segmented by detecting distinct switching points separating ARS and commuting modes. We used the Penalized Contrast Method~\citep{barraquand2008animal} -- a non-parametric method in which the initial number of segments is unknown and estimated by minimizing a penalized contrast function. First passage time (FPT) was used as the focal metric~\citep{fauchald2003using, barraquand2008animal}. Each point was assigned an FPT outside a radius of $R_s$ and data was segmented such that points with similar FPT that were close in time were clustered together~\citep{lavielle2005using}. Data are then split into ARS (within-patch movement with high FPT) and commuting (between-patch movement with low FPT) according to a threshold on the mean FPT chosen in accordance with the animal's velocity, which during commuting is $7-10$ m/s for all three species. In our segmentation, we choose $R_s = 100$ m and a threshold of $50$ s. Yet, our results are insensitive to small changes in these parameters: $R_s$ was tested between $70 - 150$ m and the FPT threshold between $40$--$120$ s [see Appendix~\ref{appendix:segmentation} and Fig. S1 in the Supplementary Information (SI)]. Note that the choice of threshold reflects the time it takes the bird to cross the area defined by $R_s$. For instance, taking $R_s = 100$ m, we defined the threshold for a commuting flights such that crossing a diameter of $200$ m takes $ < 50$ s, as if the bird flies in a velocity of $4$ m/s across in a straight line. 

Having segmented the data into commuting and ARS phases, we related the data to the CTRW model by obtaining $\alpha$ [see Eqs.~\eqref{WT_PDF} and \eqref{averagedtasd_def}] for each individual. To this end, we computed the average TASD, $\average{\tasd}$, for each bird in the following way: for each ARS, we computed the TASD and then averaged over all TASDs with similar total times $t$. Here, different times $t$ were obtained by analyzing the TASD from the time we detected the switching point to time $t$, for different values of $t$. Here,  we averaged only TASDs within the same period, either within or after the breeding season (see above).
The averaged TASD was fitted to Eq.~\eqref{averagedtasd_def}, in order to obtain the value of $\alpha$ for each individual. The fit was performed in the range of $1 < \Delta < 30$ min as to ensure the validity of the theory and that we were above the noise limit. To estimate the error around $\alpha$, it was fitted for different times $t$, and the error was taken to be a 95\% confidence interval in the slope variations for different measurement times.
Besides measuring $\alpha$ for all individuals, we also computed the mean value of $\alpha$ for different subgroups (e.g., sex and age) within each species 
\footnote{We used an unpaired t-test to compare the $\alpha$ values  between species. After verifying variance homogeneity (Levene test) and normal distribution (Shapiro-Wilk test) we performed for each two species an unpaired t-test for the means of two independent samples, with equal means as the null hypothesis. P-values were corrected for multiple comparisons using Tukey's honest significance test.}.

To directly calculate the WTs during ARS, we used a spatiotemporal criterion for the segmentation procedure with a threshold $R_{th}$. We chose the threshold to be above the noise limit of $5$--$10$ m and much smaller than the typical size of an ARS patch; the results do not vary significantly when $R_{th}$ is between $10$ and $25$ m. We fitted the observed WTs to a power-law distribution [Eq.~\eqref{WT_PDF}] using the method of maximum likelihood~\citep{clauset2009power}.
To test the quality of fit, we used a likelihood ratio test to compare between a power law and exponential fit for the distribution~\citep{clauset2009power, alstott2014powerlaw}. In all cases, a power law was a better fit. A similar test between a power law and a truncated power law showed that the latter was a better fit.

\begin{figure}[t]
	\center
	\includegraphics[width=0.95\linewidth]{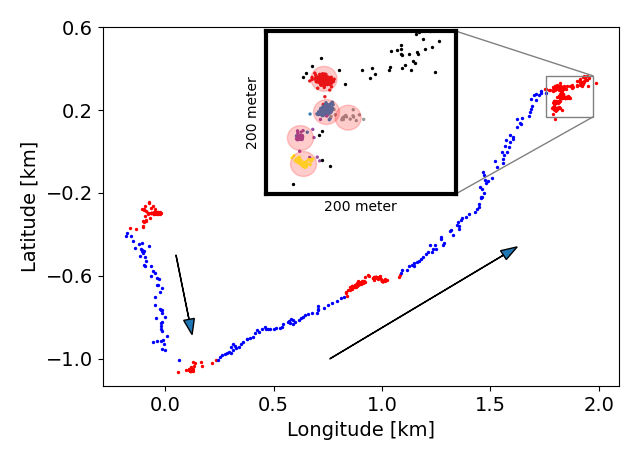}
	\vspace{-4mm}
	\caption{Foraging tracks are composed of superdiffusive commuting flights and subdiffusive local search. (a) Seventy-five minutes of tracking at 0.5 Hz (2,280 localizations) of a randomly selected female owl (tag 4782), segmented into 4 ARS (red) of 4, 2, 4 and 57 min long, and 3 commuting segments (blue), each lasts approximately 1 min. The arrows show the direction of motion. The inset shows local search within a single ARS (size $200 \times 200\, \text{m}^2$, 57 min long), with 5 local clusters (WTs of 14, 36, 2, 3, and 1 min), each represented by a different color and highlighted by a red circle. The owl performs local jumps of $25$ to $45$ m between the local clusters.
	}
	\label{fig1}
\end{figure}

\section{RESULTS}

ATLAS provided the means to obtain high-resolution trajectories, per individual, for months at a time, at a mostly constant frequency. This allows us to reveal that commuting between ARS patches (Fig.~\ref{fig1}) is qualitatively different from moving (between the more local clusters) within ARS (Fig.~\ref{fig1}, inset). Such local clusters within ARS can be located only a few tens of meters from one another, each with a different WT.
Notably, in our study resource patches that define ARS cannot be defined \textit{a-priori}, as in general they do not necessarily correspond to well-defined, discrete spatial units~\citep{benhamou2014scales}. These are thus defined via the segmentation procedure and are robust with respect to changes in the segmentation parameters (see Appendix \ref{appendix:segmentation} and Fig. S1 in the SI). 

\begin{figure}[t]
	\center
	\includegraphics[width=0.9\linewidth]{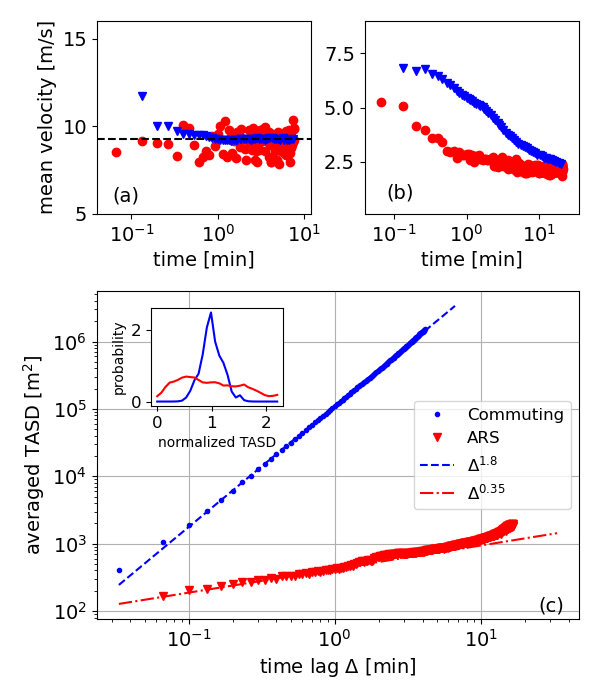}
	\vspace{-8mm}
	\caption{(a-b): The ensemble-averaged (red circles) and time-averaged (blue triangles) velocities of empirical movement tracks of a kite during commuting (a) and ARS (b). (c) The empirical averaged TASD [Eq.~\eqref{averagedtasd_def}] of ARS segments (red triangles) and commuting segments (blue dots). The dashed and dash-dotted lines are power laws with $\Delta^{0.35}$ and $\Delta^{1.80}$. 
	Inset shows the probability density of TASD normalized by its average at time lag $\Delta = 0.33\ll t$ (see text), for both cases. 
    }
	\label{fig2}
\end{figure}

Direct evidence of the ergodic and nonergodic nature of commuting and ARS respectively, is given in Fig.~\ref{fig2} by comparing the time-averaged and ensemble-averaged velocities for the two ensembles. Here, the ensemble-averaged velocity is calculated by directly averaging over an ensemble of flights, and the time-averaged velocity by averaging over time for different measurement times $t$ and then averaging over the ensemble [as in Eq.~\eqref{averagedtasd_def}]. Commuting segments (Fig.~\ref{fig2}a) are ergodic (\textit{e.g.}, for $t>1$ min the time-averaged velocity is $9.26 \pm 0.05$ m/s and the ensemble-averaged velocity is $9.10 \pm 0.55$ m/s), and show no ageing. In contrast, ARS segements (Fig.~\ref{fig2}b) are non-ergodic (\textit{e.g.}, for $t= 1$ min the time-averaged velocity is $5.41$ m/s and the ensemble-averaged velocity is $3.04$ m/s) and display ageing.
We further compare the measured diffusivity of commuting and ARS segments by calculating the averaged TASD in both cases for a representative individual female owl (Fig.~\ref{fig2}c).
The commuting segments show an averaged TASD that resembles ballistic motion, $\average{\tasd}\sim \Delta^{1.8}$. Indeed, the long commuting flights taken by owls are directed and relatively fast ($8.5$--$9.5$ m/s), and we have checked that similar fast directional commuting also holds for kites and kestrels. In contrast, the averaged TASD of ARS segments is qualitatively different and is subdiffusive, $\average{\tasd}\sim \Delta^{0.35}$. Moreover, we observe a wide distribution of TASD around the averaged TASD, characteristic of ergodicity breaking, in contrast to the sharply-peaked distribution for the commuting phase (Fig.~\ref{fig2}c inset).

\begin{figure}[t]
	\center
	\includegraphics[width=0.95 \linewidth]{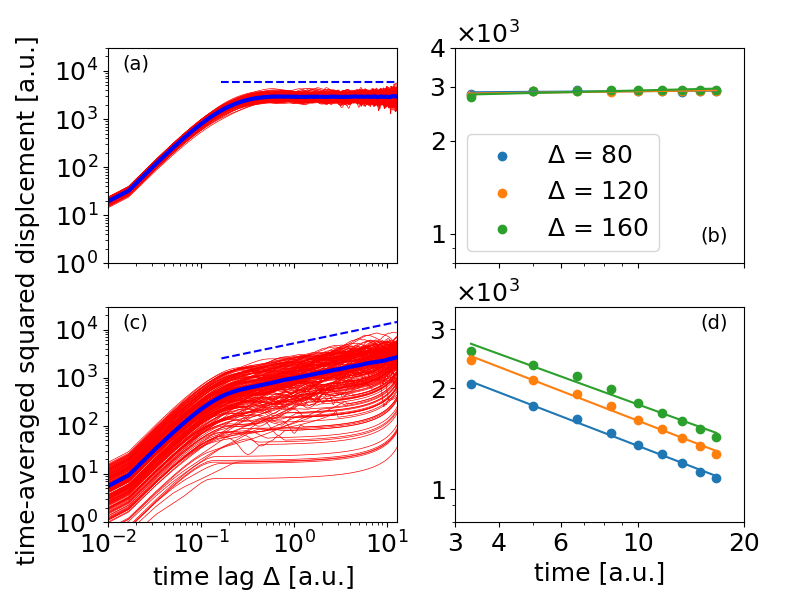}
	\vspace{-1mm}
	\caption{CTRW simulations. (a-b) Exponentially-distributed WTs (a simple RW with $\tau_0 = 10 s$). (c-d) Power law WTs, \eqref{WT_PDF}, with $\alpha = 0.6$. In (a,c) each of the red lines are the TASD calculated for a single simulation versus the time difference $\Delta$, and the blue line is the average TASD, Eq.~\eqref{averagedtasd_def}, over an ensemble of 200 simulations. The blue dashed lines are the theoretical scaling. In (b,d) we plot the averaged TASD, Eq.~\eqref{averagedtasd_def}, versus the total time of the simulation. Simulations were done in a bounded domain of $100 \times 100$.  }
	\label{fig3}
\end{figure}

To get a better understanding of the nonergodic and subdiffusive nature of movement within ARS we employ the CTRW formalism (see Sec.~\ref{sec:CTRW}).
As a first step, CTRW simulations were performed to illustrate the markedly different behavior between subdiffusive CTRW and simple RW, in the case of a confined movement within predefined domain walls, see Fig.~\ref{fig3}. For exponentially-distributed WTs, the TASD saturates upon interacting with the boundaries and there is no dependence on the measurement time $t$ (Fig.~\ref{fig3}a-b). In contrast, for power-law-distributed WTs, the averaged TASD does not saturate with time lag $\Delta$ and the dependence on $t$ agrees with Eq.~\eqref{averagedtasd_def} (Fig.~\ref{fig3}c-d). Here, as very long WTs can occur, time-averaged measurable quantities are generally irreproducible, such that realizations vary from one another even at very long times~\citep{metzler2014anomalous}, see Fig.~\ref{fig3}c.

\begin{figure}[t]
	\center
	\includegraphics[width=0.9\linewidth]{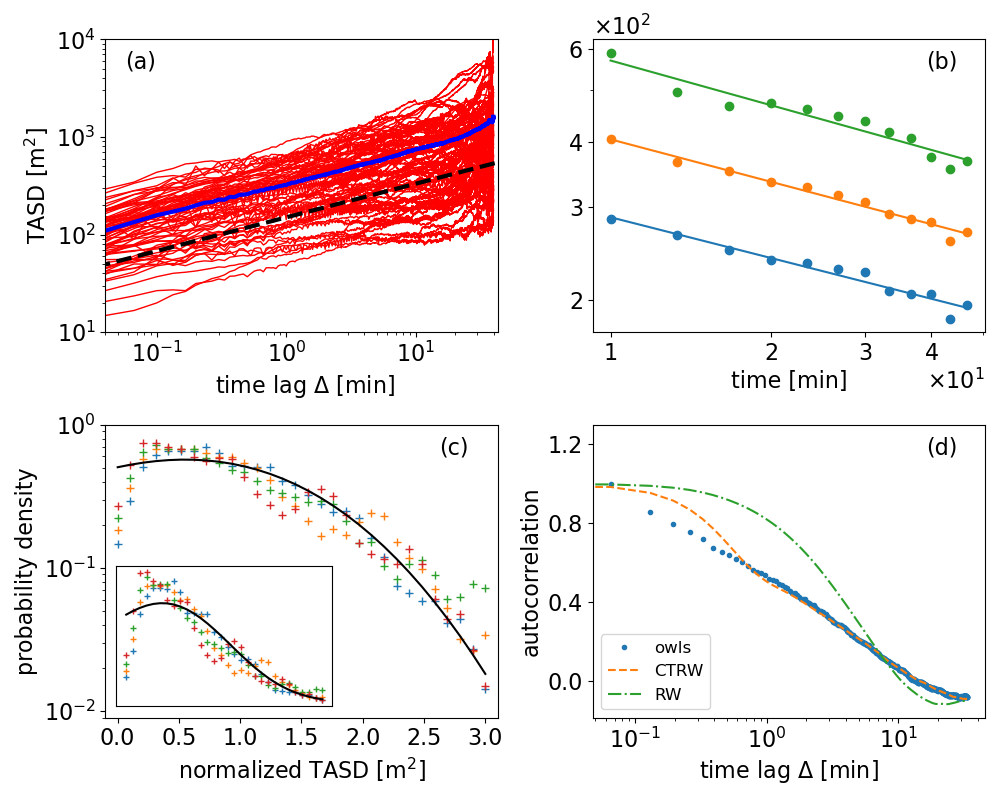}
	\vspace{-6mm}
	\caption{(a) TASD (red solid lines) and averaged TASD (blue solid line) of 40-min ARSs of an owl, over time lag $\Delta$. The dashed line scales as $\Delta^{0.4}$. (b) The averaged TASD over time $t$ for $\Delta$=10 s (blue), 40 s (orange) and 120 s (green), yielding powers of $-0.26, -0.27$ and $-0.28$ respectively.
	(c) The probability density of the normalized TASD [$\phi(\xi)$] during ARS for the same individual for $\Delta = 100, 200, 400, 700$ s. Dashed line is the  Mittag-Leffler distribution [Eq.~\eqref{phi_xi}] for $\alpha = 0.6$. Inset shows the same plot on a linear scale.  (d) The autocorrelation function (blue dots)
	averaged over 100 ARS segments of $2000 s$. The orange dashed and the green dash-dotted lines are averages over 1000 simulations of subdiffusive CTRW (with $\alpha = 0.6$) and simple RW, respectively.}
	\label{fig4}
\end{figure}

Next, we plotted the averaged TASD of another randomly selected breeding female owl versus the time lag $\Delta$ on a log-log scale, revealing a slope approaching a value of $0.4 \pm 0.05$ (Fig.~\ref{fig4}a, compare Fig.~\ref{fig3}c). Here, the notable spread of TASD for single trajectories around the average TASD [Eq.~\eqref{averagedtasd_def}] implies that different ARS segments are highly variable even for the same individual owl which indicates ergodicity breaking (compare Fig.~\ref{fig3}c).
In addition, the averaged TASD versus $t$, for different values of $\Delta$, shows an explicit dependence on the measurement time (Fig.~\ref{fig4}b, compare Fig.~\ref{fig3}d), which is characteristic of ageing, as longer measurement times infer smaller measured diffusivity~\citep{metzler2014anomalous}. Thus, our analysis (Fig.~\ref{fig4}a-b) indicates that the process is nonergodic and subdiffusive. Moreover, the dependence $\average\tasd \sim \Delta^{0.4}$ indicates that for this individual, $\alpha \simeq 0.6$. Note that, in our segmentation procedure there is uncertainty in the onset of ARS, which brings about uncertainty in the initial measurement time (ageing). As a result, we extract the $\alpha$ value from the dependence of the averaged TASD on $\Delta$, which we found in simulations to be unaffected by ageing, and \textit{not} from the dependence on $t$, which clearly depends on ageing~\citep{metzler2014anomalous}, see SI, Fig. S2.

Having recorded the value of $\alpha$ for each individual bird, the mean value of $\alpha$ for different subgroups within each species is given in Table 1.
Within each subgroup the variation in $\alpha$ between individuals is found to be small (typically smaller or similar to the measurement error in the measurements of $\alpha$). Moreover, we found that within each species, the values of $\alpha$ do not statistically differ between males and females, fledglings and adults, and during or after the breeding season. In addition, we found no difference between kestrels and kites (p-value $> 0.05$), yet owls had significantly larger values of $\alpha$ than either kites or kestrels (p-value $=0.001$ for both). Comparing adults and fledglings yields no significant differences for owls and kites. Comparing breeding and post-breeding periods for owls (for kites and kestrels we did not have sufficient data outside the breeding season to make such a comparison), we found higher  $\alpha$ values  during breeding for all 14 individual owls, but the null hypothesis of identical averages was not rejected (p-value $=0.1$).

The variability of the TASD (the spread of individual TASDs around their average) is quantified in terms of the dimensionless parameter $ \xi = \tasd/\average{\tasd}$. At long measurement times the distribution of $\xi $ satisfies a Mittag-Leffler distribution~\citep{aaronson1997introduction, he2008random, burov2010aging}
\begin{equation} \label{phi_xi}
\phi(\xi) = \frac{\Gamma^{1/\alpha}(1 + \alpha)}{\alpha \xi^{1 + 1/\alpha}} l_\alpha\left( \frac{\Gamma^{1/\alpha}(1 + \alpha)}{\xi^{1/\alpha}}\right),
\end{equation}
which also holds for subdiffusive CTRW in a bounded domain~\citep{burov2010aging,neusius2009subdiffusion}. Here, $l_\alpha$ is the one-sided L\'evy stable distribution with the Laplace transform $\mathcal{L}\{l_\alpha (t)\} = \exp(-u^\alpha)$, while $\Gamma(\cdot)$ is the Gamma function~\footnote{The Aaronson-Darlin-Kac theorem predicts that the distribution of time averages of a process with an infinite measure will be given by the Mittag-Leffler distribution~\citep{aaronson1997introduction, aghion2020infinite}.}.
For Brownian diffusion, $\alpha \to 1$, $\phi(\xi) \sim \delta(\xi - 1)$, a sharply peaked distribution around 1. However, for general $\alpha$ the distribution is wide and skewed; \textit{e.g.}, for $\alpha = 0.5$, $\phi(\xi)$ tends to a half Gaussian with maximum at $\xi = 0$. 

To evaluate the reproducibility of individual ARS, we calculated the distribution $\phi(\xi)$ for different $\Delta$ for each individual, theoretically predicted by Eq.~\eqref{phi_xi}. In Fig.~\ref{fig4}c we compare between theoretical  (for $\alpha=0.6$) and empirical results for one female owl (see also Fig.~\ref{fig2}c inset). Importantly, this broad distribution, which does not seem to depend on $\Delta$, is observed for all individuals and serves as further evidence of ergodicity breaking.
The fact that the empirical $\phi(\xi)$ is more sharply peaked than the theoretical prediction, can be explained by the presence of noise in our data, which skews the $\xi$ distribution and yields lower than expected values close to $\xi = 0$~\citep{jeon2013noisy}. To provide yet another verification of the CTRW model, we calculated the autocorrelation function $R_{XX}(\Delta) = E\left[ X(t)X(t+\Delta)\right]$, versus the time lag $\Delta$, for owls averaged over many ARS segments (Fig.~\ref{fig4}d), where $X$ is the bird's location normalized by its mean.
This reveals that simulations based on subdiffusive CTRW fit the data much better than those based on simple RW. This is also confirmed independently by a p-variation test~\citep{magdziarz2009fractional} (see Appendix \ref{appendix:p} and Fig. S3 in the SI).

\begin{table}[t]
\label{tab:alphavalues}
\centering
\begin{tabular}{lllll}
\toprule\toprule
species   & subgroup            & group size & $\left<\alpha\right>$ & $\pm 95\%$ \\ \midrule
owls      & breeding adults      & 14 & 0.69 \; & 0.05    \\
          & post-breeding adults & 14 & 0.65 \;  & 0.05    \\
          & fledglings           & 30 & 0.66 \;  & 0.03    \\
kites     & adults               & 6  & 0.56 \;  & 0.05    \\
          & fledglings           & 10 & 0.51  \; & 0.05    \\
kestrels  & adults               & 10 & 0.51 \;  & 0.09    \\ \bottomrule\bottomrule
\end{tabular}
\caption{\textmd{Mean $\alpha$ values for species subgroups, including the individuals number in each subgroup and a $95\%$ confidence interval around the mean. Subgroups are mutually exclusive except adult owls with tracks of the same 14 birds divided to breeding and post-breeding periods.} }
\end{table}

\begin{figure}[t]
\centering
\hspace{-5mm}
\includegraphics[width=1.04\linewidth]{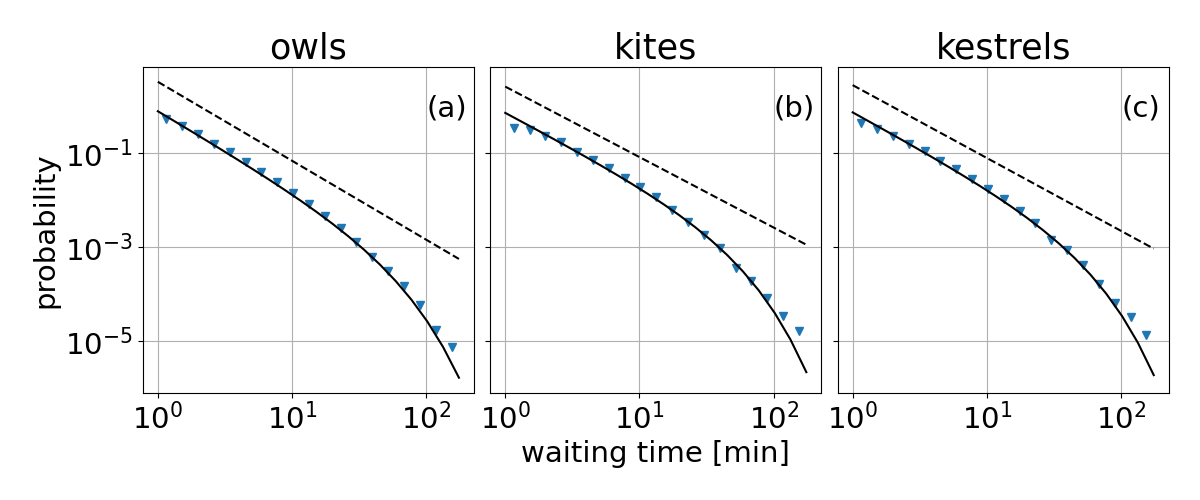}
\vspace{-7mm}
	\caption{WT probability densities (symbols) within a radius of $15$ m, for 14 owls (a), 6 kites (b) and 10 kestrels (c), all adults. The solid line is a fit to a truncated power law, $P(\tau) \sim \tau^{-1-\alpha} e^{-\tau/\tau_0}$, with $\alpha = 0.68, 0.51, 0.55$, and $\tau_0 = 40, 35, 36$ in (a), (b) and (c), respectively. These values of $\alpha$ are the average fit values for the joint distribution of all birds within each species, and the power law truncates at $\tau_0 = 20$--$80 $ min for all 70 individual birds. The dashed lines are power laws, [Eq.~\eqref{WT_PDF}], with the same $\alpha$ values. 
	}
	\label{fig5}
\end{figure}

Based on the subdiffusive CTRW model presented above, the observed movement patterns are a result of long WTs. We show direct and independent evidence for long WTs by calculating the distribution of WTs within a radius of $15$ m (chosen to be significantly smaller than the typical ARS radius but larger than the localization error, see Sec. \ref{Sec:data}), about the size of a local cluster within ARS (Fig.~\ref{fig1}, inset), for all three study species. In Fig.~\ref{fig5} we plot the WT distribution of all individuals within each species, while similar results were separately obtained for each individual (SI, Fig. S4). For most individual birds, the results of this fit were within the error of the value of $\alpha$, independently found by fitting the averaged TASD to Eq.~\eqref{averagedtasd_def} (see Table 1) and are thus an independent validation of our proposed model. Importantly, the distribution of WTs follows a power law with a diverging mean, which however, breaks down at an average time of $40$ min, as was corroborated by fitting a truncated power law to the data (Fig.~\ref{fig5})~\footnote{Note that, the analysis presented in Fig. \ref{fig4} is also limited to approximately 40 min (depending on the individual), as very few ARS segments last longer than that. }.

\begin{figure}[t]
\centering
\hspace{-4mm}\includegraphics[width=1.04\linewidth]{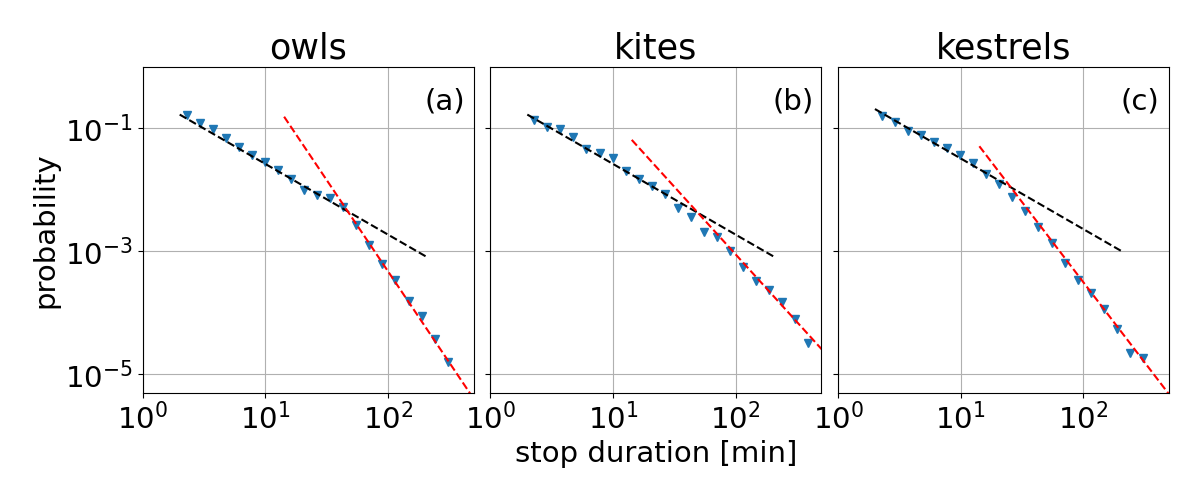}
\vspace{-6mm}
	\caption{Distributions of stop durations $\Delta T$ (symbols), defined as the total time spent within single ARSs (see text), for owls (a), kites (b) and kestrels (c), all adults. The dashed black and red lines are power laws, presenting a behavioral switch from $\Delta T^{1.15}$ for shorter $\Delta T$ for all three species to $\Delta T^{2.97}, \Delta T^{2.20}$ and $\Delta T^{2.61}$ for longer $\Delta T$, with typical transition times (when the two slopes intersect) of $\Delta T = 45, 48$ and $36$ min in (a), (b) and (c), respectively.}
	\label{fig6}
\end{figure}

To better understand this regime shift at 40 min, we compute the \textit{total time} spent in a single ARS -- the \textit{stop duration} $\Delta T$  (to be distinguished from the WTs in Fig.~\ref{fig5}). Here, we find a behavioral switch at 30--60 min for all three species (Fig.~\ref{fig6}) \footnote{We note that the regime shift is inferred primarily from the detailed analysis of Fig. \ref{fig5}, where Fig.~\ref{fig6} is a summary statistics that enables us to interpret this result.}. For short times ($< 40$ min), the distribution of stop durations is best fitted by a power-law $\Delta T^{\gamma}$ with $\gamma = 1.15$, whereas for longer times ($>40$ min), the data is best fitted by a power-law with $\gamma > 2$ for all species.
These results are insensitive to our specific definition of ARS, see Appendix~\ref{appendix:segmentation}. 
This distribution of $\Delta T$ indicates that the birds abruptly shift from a subdiffusive to a (super)diffusive regime as stops longer than $40$ min become exceedingly unlikely (see  Appendix \ref{Appendix:WT}).
We thus suggest that the long WTs within ARS, driven by the motivation to hunt from a perch, to rest, or by other reasons, are not only spatially but also temporally confined, such that beyond $\sim 40$ min the functional gain from a long stop is diminished, driving the bird to move to another location outside the local ARS patch.

\section{DISCUSSION}

Applying the CTRW framework to rich high-quality movement datasets encompassing $>\!7\times 10^7$ localizations from 70 individuals of three avian predator species, we revealed that local ARS is uniquely characterized as nonergodic, irreproducible and subdiffusive, whereas commuting is ergodic, reproducible and superdiffusive. 
Here, ARS combines short and very long WTs yielding nonergodic motion, while  commuting  includes long directional flights, and is ergodic as implied by the TASD distribution around the mean. These two distinct modes are separated via a behavioral switch in animal movement occurring at characteristic spatial and temporal scales. Importantly, we \textit{quantified} these temporal (below and above 40 min) and spatial  (ARS and commuting) transition scales, thus providing new insights into the well known problem of ``pattern and scale in ecology" \cite{levin1992problem, chave2013problem}. Our analysis also shows that notions of universal foraging behavior and scale-free movement should be replaced by case- and scale-specific behavior and movement~\cite{viswanathan2011physics,reynolds2015liberating}.  

In previous studies that did not assess ergodicity, the \textit{average} WT during ARS has been used to distinguish foraging tactics among species (see, \textit{e.g.},~\cite{halperin2018use}). In contrast, our study shows that the \textit{distribution} of WTs during search is fat-tailed, with a cutoff around 40 min, indicating that ARS is composed of multiple foraging tactics. Moreover, whereas ARS is characterized by ageing, commuting birds maintain similar movement characteristics regardless of commuting duration. Putting this in simple words, our study reveals that there are many ways to hunt within a local patch but only a limited number of ways to commute between distant patches.

Within species, $\alpha$ values were similar for both adults and fledglings and were only slightly higher for adult owls during breeding, compared to post-breeding (but within the same statistical error). 
The tendency of breeding owls to minimize WTs within ARS is expected due to the urge to provision their nesting mate or nestlings. The lack of a significant difference between breeding and post-breeding owls can be attributed to the urge of adult owls to provision their young also a few months after they fledge. The only significant difference we found is among species: while very long for all three species, WTs of kestrels and kites were similar and longer (lower $\alpha$) than those of owls, suggesting that the nocturnal owls tend to remain stationary for shorter periods during foraging compared to the two diurnal raptors. 

What is the origin of these slow, subdiffusive, and nonergodic dynamics within ARS? Although slow dynamics are counter-intuitive, they make sense if we recall that the predators primarily forage to find their target rather than to cover the most ground. We postulate that hunting efficiency can increase via highly variable, irreproducible foraging tactics compared to less variable, reproducible ones. Such variability may hold an evolutionary advantage through higher individual fitness of avian predators that combine different foraging behaviors~\citep{bartumeus2016foraging}. 
We suggest that long WTs and ergodicity breaking are evolutionarily coupled to prey behavior. Barn owls, for example, are well known for their acute night vision and their high auditory sensitivity enabling high spatial resolution in sound localization of their prey even in complete darkness~\citep{payne1971acoustic, konishi1973owl}. Their prey, however, have evolved to avoid owl predation using various strategies such as minimizing exposure in risky times and habitats~\citep{abramsky1996effect} and adopting escape strategies during an active owl attack~\citep{edut2004protean, ilany2008wait}. Controlled experiments in a closed arena revealed that owls tend to postpone their attack until their prey became motionless~\citep{ilany2008wait}, and there is a high variation in capture duration (from first attack to a successful capture) ranging from 0.5 sec to 43 min~\citep{edut2004protean}. Thus, as it is hard to catch a highly apprehensive moving prey, adopting irreproducible (and thus unpredictable) movement tactics may prove beneficial for a predator, 
rather than committing to a single tactic or behavior.

In summary, our analysis offers novel insights and a general formalism for better quantifying scale-specific behavior at different spatiotemporal scales~\citep{benhamou2014scales, riotte2020environmental}.
As long WTs have also been  observed in other species~\cite{wearmouth2014scaling, bartumeus2010fishery, bazazi2012intermittent, zhao2016understanding, brockmann2006scaling}, we expect our formalism to be highly relevant for analyzing scale-specific behavior in these systems as well.
More broadly, the lack of a typical behavior during ARS highlights the importance of considering ergodicity in movement ecology research, in general. Indeed, the ergodicity assumption can either lead to  over- or under-estimation of within-patch movement, depending on the averaging method (time averaging or  ensemble averaging, respectively).
These results also have potential implications on the fundamental issue of estimating population-level traits from observed movement patterns \cite{zeller2012estimating, ovaskainen2019joint}. For instance, most spatial capture-recapture models \cite{efford2004density, fleming2021experimental} used for estimating population-level encounters and landscape resistant assume that expected point locations are static variables that can be averages over an ensemble (an individual's location may change, but the expected behavior, averaged over an ensemble, does not) \cite{royle2018unifying}. As this assumption is not valid for nonergodic systems, inferring population-level dynamics from individual movement for such systems is highly nontrivial, and should be at the center of future work. Moreover, nonergodic movement patterns, which may lead to non-exponential inter-individual interaction frequencies, are expected to affect competition between individuals \cite{bonyah2019fractional}.

Beyond movement ecology, we stress that considering ergodicity is key in a wide range of dynamical systems and stochastic processes in ecology, evolutionary biology and behavioral sciences~\citep{mangalam2021point}. For example, in studies of the speed-accuracy trade-off in cognitive psychology, averaging individual behavior over time versus averaging over an ensemble of individuals can yield opposing results, a phenomena equivalent to \textit{Simpson's paradox}~\citep{mangalam2021point}. 
Since these and many other scenarios can lead to ergodicity breaking, assessment of ergodicity should become a routine practice in these fields of research.    


\section{ACKNOWLEDGEMENTS}
For fieldwork and technical assistance we thank Y. Bartan, A. Levi, S. Margalit, R. Shaish, G. Rozman and other members of the Movement Ecology Lab and the Minerva Center for Movement Ecology. We also thank R. Metzler for useful comments.  O.V. and M.A. acknowledge support from the ISF grant 531/20. ATLAS development, maintenance, and studies have been supported by the Minerva Center for Movement Ecology, the Minerva Foundation, and ISF grants 965/15 and 1919/19 to R.N and S.T.; research on black-winged kites was supported also by JNF/KKL grant 60-01-221-18. R.N. also acknowledges support from Adelina and Massimo Della Pergola Chair of Life Sciences.

\section{SUPPORTING INFORMATION}
Additional Supporting Information may be downloaded via the online version of this article.

\appendix

\section{Mathematical background} \label{appendix:math}
In this appendix we give further details on continuous-time random walks (CTRWs) and ergodicity breaking, and provide a detailed derivation of Eq.~\eqref{averagedtasd_def}. We also discuss the universality of this result with respect to the confining potential.

A CTRW is a generalization of a random walk, where a diffusing particle has to wait  a random waiting time $\tau$ before making a local jump of length $\delta r$, where  $\tau$ and $\delta r$ are drawn from probability distributions $\psi(\tau)$ and $\phi(\delta r)$ respectively~\cite{metzler2014anomalous}. Here, e.g., Pearson's random walk~\cite{pearson1905problem} in discrete space is retrieved when $\psi(\tau) = \tau_0^{-1} \exp(-\tau/\tau_0)$ and $\phi(\delta r) = \delta_{\delta r, 1}$, where $\tau_0$ is the mean WT and $\delta_{i, j}$ is the Kronecker delta. In the limit of long times in the CTRW model, and assuming that $\psi(\tau)$ given by Eq. \eqref{WT_PDF} with $0<\alpha <1$, one can write a (one-dimensional) fractional Fokker-Planck equation governing the probability density $W(x, t)$ of being at position $x$ at time $t$~\cite{metzler2000random}:
\begin{equation} \label{FFPE}
\frac{\pp}{\pp t} W(x, t) = K_\alpha \;\; \\_0 D_t^{1-\alpha} \left\{-\pd{}{x} \left[\frac{F(x)}{k_B T}\right] + \pdd{}{x}\right\}\; W(x, t).
\end{equation}
Here $V(x) = -\int^x F(x')dx'$ is the confining potential of the random walker, while for a free (unconfined) walker, $F(x)=0$.
Furthermore, $K_{\alpha}$ is a generalized diffusion parameter, and the Riemann Liouville operator  $_0 D_t^{1-\alpha} \equiv (\pp/\pp t) {_0} D_t^{-\alpha}$  is defined for $0< \alpha < 1$ as~\cite{oldham1974fractional}:
\begin{equation}
_0 D_t^{1-\alpha} \phi(x, t) \equiv \frac{1}{\Gamma(\alpha)}\pd{ }{t} \int_0^t dt' \frac{\phi(x, t')}{(t - t')^{1-\alpha}}.
\end{equation}
In the limit of $\alpha \to 1$, Eq.~\eqref{FFPE} reduces to the Fokker-Planck equation as the Riemann Liouville operator reduces to the unity operator.
For an unconfined random walker the MSD and averaged TASD [defined by Eq.~\eqref{averagedtasd_def}], are respectively given by~\cite{metzler2000random,he2008random}
\begin{eqnarray}\label{msdtasd}
\average{x^2} \sim t^{\alpha}\,,\;\;\;\;\; \average{\tasd}\sim \frac{\Delta}{t^{1 - \alpha}}.
\end{eqnarray}
As the dependence of the averaged TASD on $\Delta$ is different from the dependence of the MSD on the measurement time $t$, the process is said to display weak ergodicity breaking (to be distinguished from \textit{strong} ergodicity breaking where phase space is separated into non-accessible domains). Moreover, the explicit dependence of the averaged TASD on  $t$ indicates an ageing effect~\cite{metzler2014anomalous}.

While for Brownian motion and most ergodic processes the TASD saturates upon interacting with the confinement, for  subdiffusive CTRW in a bounded domain, $F(x)\neq 0$, governed by Eq.~\eqref{FFPE}, the averaged TASD does not saturate. Defining  $\average{x^n}_B= \mathcal{Z}^{-1} \int_{\infty}^{\infty} x^n \exp\left( - V(x)/k_B T\right)$ as the $n^{\text{th}}$ moment of the Boltzmann distribution,
with a normalizing factor of $\mathcal{Z} = \int_{\infty}^{\infty} \exp\left( - V(x)/k_B T\right)$, one obtains for $1 \ll \Delta \ll t$~\cite{burov2010aging}:
\begin{equation} \label{averagedtasd}
\average{\tasd} \sim \left( \average{x^2}_B - \average{x}_B^2\right)\left( \frac{\Delta}{t}\right)^{1-\alpha}.
\end{equation}
Here, the dependence of the averaged TASD on $\Delta$ for unbounded and bounded CTRW  [Eqs.~(\ref{msdtasd}) and (\ref{averagedtasd}), respectively] is markedly different for $\alpha>0$.
We stress that Eq.~\eqref{averagedtasd} is universal, and in the leading order, does not depend on the confining potential. Indeed, the potential only enters in the prefactor, and via the first two moments of the distribution~\cite{burov2010aging}.
Therefore, knowledge of the exact form of the confining potential of individual birds is \textit{not} crucial for any of our  conclusions.

Finally, in the simulations presented in Fig.~\ref{fig3}, the jump lengths were chosen such that at each time step the random walker can move to any point within the predefined domain walls with equal probability. As in CTRW, generally the jump length $\delta r$ is a random variable that is drawn from a jump-length distribution $\phi(\delta r)$, we verified using simulations that our conclusions, and in particular Eq.~(\ref{averagedtasd_def}), do not depend on the specific choice of $\phi(\delta r)$.



\section{Segmentation procedure} \label{appendix:segmentation}
Here, we provide evidence that our main results do not depend on the segmentation method.
In the Penalized Contrast Method we used $R_s = 100$ m and a threshold of $\tau_s = 50$ s to classify commuting flights from ARS. 
In Fig.~S1a (SI) we varied $R_s$ between $70$--$150$ m to show that the value of $\alpha$ converges at around  $R_s = 100$ m to a constant value. In Fig.~S1b (SI) we varied $\tau_s$ between $40$--$200$ s to show that the value of $\alpha$ does not depend on this parameter. Similar sensitivity analysis was performed for all individuals.  Notably, due to the high frequency and resolution of the data the segments are visible to the eye, see Fig.~\ref{fig1}, and are not sensitive to any specific segmentation procedure.  To further show this, switching points were also detected using spatiotemporal criteria segmentation, such that localizations that are in proximity to one another, both in space and time were segmented together~\cite{gurarie2016animal}. Using this segmentation procedure did not significantly alter any of the results reported in our study.

\vspace{-5mm}
\section{p-variation test} \label{appendix:p}

A p-variation test was performed in order to distinguish the non-Gaussian CTRW from other types of subdiffusive behaviors such as the Gaussian fractional Brownian motion~\cite{metzler2014anomalous, magdziarz2010detecting, magdziarz2009fractional}. Notably, this test was applied in Ref.~\cite{jeon2013noisy} to evaluate the effect of noise in subdiffusive CTRW, and it was found in simulations that up to some noise level the p-variation test is valid.
The test is defined in terms of the sum of increments of a trajectory $x(t)$ on the time interval $[0, T]$:
\begin{equation}
V_n^{(p)}(t) \!=\!\! \sum_{j = 0}^{2^n\! -\! 1}  \left|  x\!\left( \!\min \left\{ \frac{(j\!+\!1)T}{2^n}, n\right\}  \!\right) \! -\! x\!\left(\!\min \left\{ \frac{j T}{2^n}, n\right\}\!  \right)  \right|^p\!.
\end{equation}
For subdiffusive CTRW, $V^{(p)}(t) = \lim_{n\to \infty} V_n^{(p)}(t)$ displays the following properties: for $p = 2$ it shows a monotonic, step-like increase in time, while for $p=2/\alpha$: $V^{(2/\alpha)}(t) = 0$~\cite{magdziarz2009fractional, magdziarz2010detecting}. In Fig.~S3 (SI) we demonstrate this test on a randomly chosen ARS segment of a female barn owl, and the results fit both theoretical predictions, indicating that the motion is a subdiffusive CTRW. We have repeated the test on many randomly chosen trajectories of various individuals, and all gave similar results.


\section{Waiting-time and jump-length distributions} \label{Appendix:WT}
Here we provide details on the WT distribution for individual birds, and on the jump-length distribution for each species. In Fig.~\ref{fig5} we plotted the WT distribution within a radius of $15$ m for each species. In the SI (Fig. S4), we plotted the WT distribution within the same radius for each individual. The power laws were fitted in a similar manner to Fig.~\ref{fig5}, see Sec. \ref{stat_ana}. A likelihood ratio test comparing between a power-law and exponential fits showed that for all individuals, a power law was at least plausible, and in most cases a better fit.
In Fig.~\ref{fig6} we plotted the distribution of the total time spent in an ARS. In the SI (Fig.~S5), we plotted the distribution of stop duration for each individual, by directly measuring the time spent in a radius of $R_{th} = 100$ m. The stop durations in Fig.~S5 were calculated using a spatiotemporal segmentation procedure 
and we verified that the results are not sensitive to small changes in $R_{th}$, between $70-200$ m. The distributions in Fig.~S5 strongly resemble those shown in Fig. \ref{fig6}.

In the SI (Fig.~S6) we further compared ARS to commuting flights by comparing the distribution of small-scale jump lengths performed \textit{within} ARS, and large-scale jump lengths of commuting flights \textit{between} ARS for owls, kites and kestrels. Both the local jumps and the commuting flights were obtained directly from our segmentation procedure, and the distributions were fitted using the method of maximum likelihood. To evaluate the distributions for each flight mode we preformed a likelihood comparison test between a power-law and an exponential distribution for each group in Fig. S6. We found that for ARS a power law was a better fit while for commuting an exponential distribution was a better fit.
Notably, for all distributions in Fig.~S6 it is plausible that other fat-tailed distributions could give a better fit; yet, we view the qualitative difference between the distributions at small and large length scales as further evidence of a qualitative switch between the intensive ARS flight mode and  extensive commuting flight mode.




\bibliography{references}

\clearpage

\setcounter{equation}{0}
\setcounter{figure}{0}
\setcounter{table}{0}
\setcounter{page}{1}
\setcounter{section}{0}
\makeatletter
\renewcommand{\theequation}{S\arabic{equation}}
\renewcommand{\thefigure}{S\arabic{figure}}
\renewcommand{\thesection}{S\arabic{section}}

\onecolumngrid
\section*{Supplementary Information}
\begin{figure}[b]
	\center
	\includegraphics[width=0.9\linewidth]{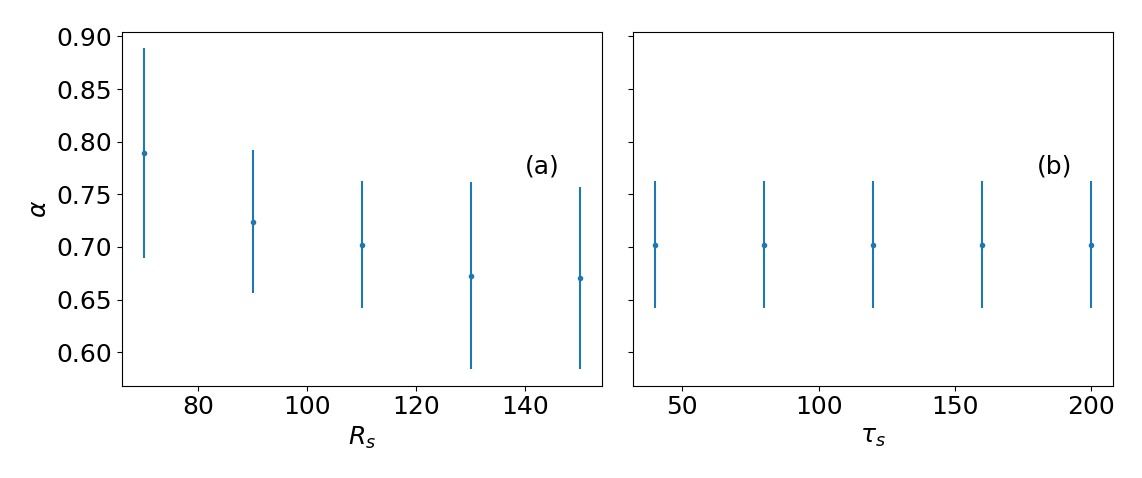}
	\caption{ (a) Value of $\alpha$ as a function of $R_s$, and (b) value of $\alpha$ as a function of $\tau_s$, for an individual owl. The error bars reflect a $95 \%$ confidence interval around the mean.}
	\label{figS1}
\end{figure}

\begin{figure}[b]
	\center
	\includegraphics[width=1\linewidth]{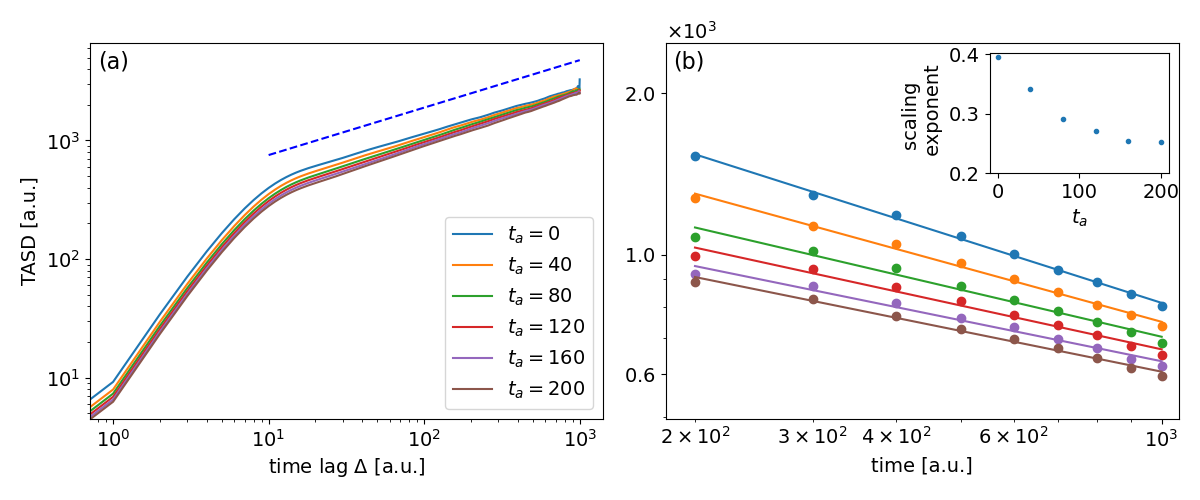}
	\caption{Ageing effects in CTRW simulations with power law WTs, see Eq. (1), with $\alpha = 0.6$. For each ageing time $t_a$, simulated trajectories are "aged" by starting at time $t_a$, i.e., all points occurring at $t< t_a$ are discarded and the TASD is calculated from the remaining points in the range $[t_a, T]$, $T = 1500$ being the simulation time. The average TASD is plotted as a function of (a) the time lag $\Delta$ and (b) the measurement time $t - t_a > 0$, where the different colors correspond to different $t_a$, see legend in panel (a).
	While the scaling of the TASD on the time-lag in (a) weakly depends on $t_a$, the scaling of the TASD on the measurement time is strongly affected by ageing, see inset.
	Here, 5000 simulations were done in a bounded domain of $100 \times 100$.}
	\label{figS5}
\end{figure}

\begin{figure}[t]
	\center
	\includegraphics[width=.9\linewidth]{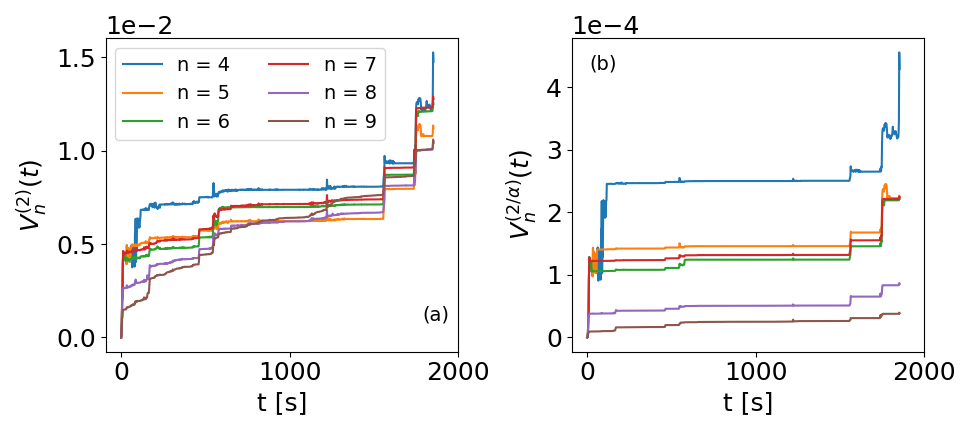}
	\caption{A p-variation test on a randomly chosen movement segment of a female owl. In (a) shown is the test for $p = 2$, and $ V_n^{(p)}(t)$ displays a monotonic step-like increase. In (b) shown is the test for $p = 2/\alpha$ for $\alpha = 0.67$ (this was the value found for this female), and as expected $V_n^{(p)}(t)$ tends to zero as $n$ is increased. Note that $n$ can only be increased up to $2^n = N$, $N$ being the number of data points in the trajectory. }
	\label{figS2}
\end{figure}

\begin{figure}[t]
	\center
	\includegraphics[width=0.9\linewidth]{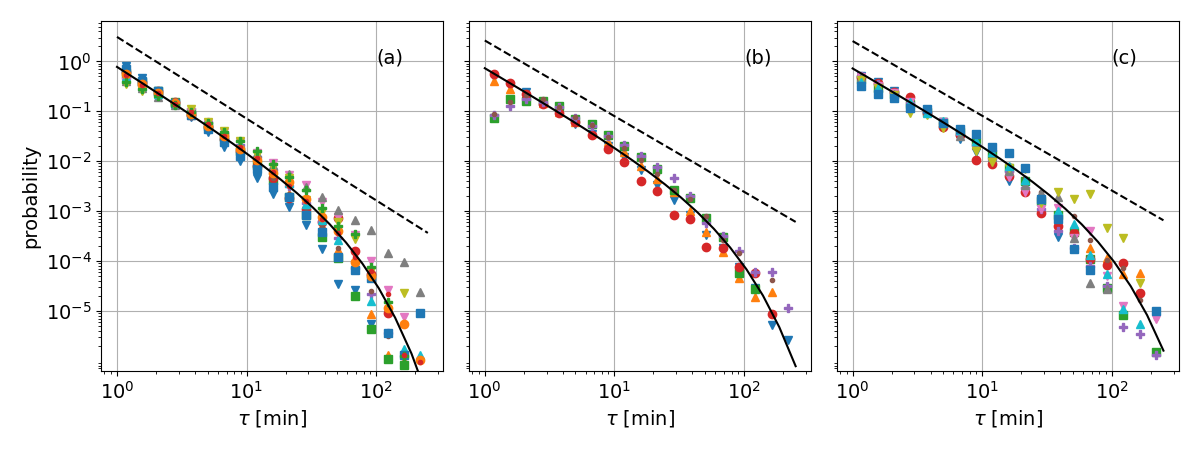}
	\caption{WTs distributions within a radius of $15$ m, for 14 adult owls (a), 6 adult kites (b) and 12 adult kestrels (c). In each panel, different shapes and colors represent different individuals. The dashed black lines indicate a power law [see Eq. (2) of the main text], with $\alpha = 0.68, 0.51$ and $0.55$ for (a), (b) and (c) respectively, and are plotted to guide the eye. The solid black line is the average fit for a truncated power law $P(\tau) \sim \tau^{-1-\alpha} e^{-\tau/\tau_0}$. For all birds the power law truncates between $\tau_0 = 20$ and $\tau_0 = 80 $ min. The average fit values, for the joint distribution of all individuals within each species, were $\alpha = 0.68, 0.51, 0.55$ and $\tau_0 = 40, 35, 36$ min for the owls, kites, and kestrels respectively. }
	\label{figS3a}
\end{figure}

\begin{figure}[t]
	\center
	\includegraphics[width=0.9\linewidth]{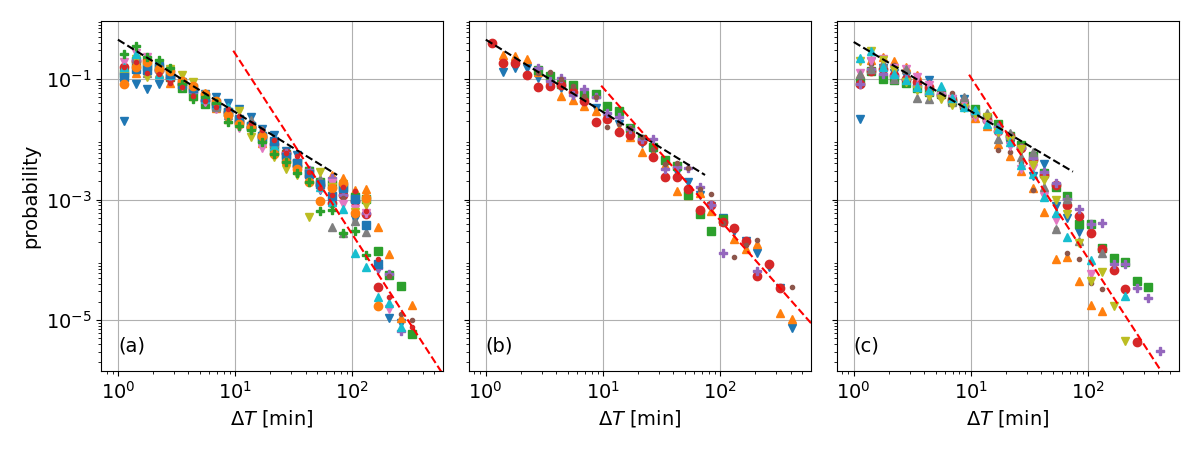}
	\caption{Distributions of time spent within a radius of $100$ m (i.e., stop duration, see main text), for 14 adult owls (a), 6 adult kites (b) and 12 adult kestrels (c). In each panel, different shapes and colors represent different individuals. The dashed black line indicates a power law with $\alpha = 1.15$ and the dashed red line indicates a power law with $\alpha = 2.97, 2.20$ and $2.61$, for the owls, kites and kestrels respectively. At short times the birds are almost stationary (see main text) indicating motion within ARS (area restricted search), while at long times the WTs display a different movement phase. }
	\label{figS3b}
\end{figure}

\begin{figure}[t]
	\center
	\includegraphics[width=0.9\linewidth]{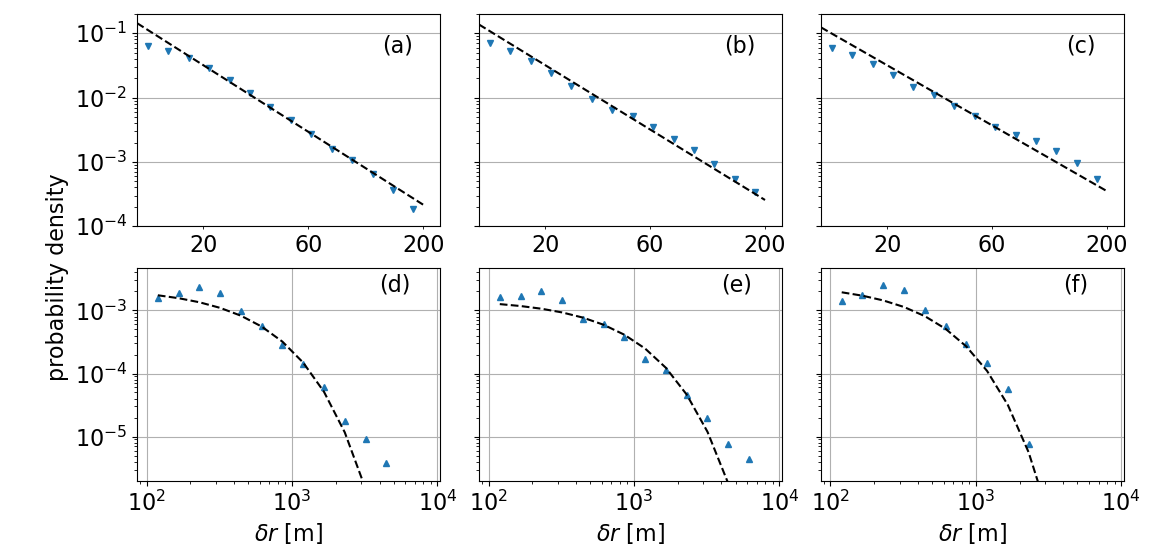}
	\caption{(a-c) Distribution of local jump length \textit{within} ARS for adult owls (a), adult kites (b) and adult kestrels (c). The dashed lines indicate a power law, $\phi(\delta r) \sim \delta r^\beta$, with $\beta = 2.17, 2.1$ and $1.96$ for (a), (b) and (c), respectively, and represent the best fit parameters. (d-f) Distribution of commuting flight distances \textit{between} ARS for adult owls (d), adult kites (e) and adult kestrels (f). The dashed  lines indicate an exponential distribution $\phi(\delta r) = e^{\delta r/\lambda}/\lambda$ with $\lambda = 437, 662$ and $374$ m for (d), (e) and (f),  respectively (see main text for details).   }
	\label{figS4}
\end{figure}

In this supplementary information we provide Figs. \ref{figS1}-\ref{figS4} to support the discussion in the main text.
In what follows, the notations and abbreviations are the same as in the main text and the equations and figures refer to those therein.

\end{document}